# Determining the Phase and Amplitude Distortion of a Wavefront using a Plenoptic Sensor

**Chensheng Wu, Jonathan Ko, Christopher C. Davis**

*Department of Electrical and Computer Engineering, University of Maryland, College Park, MD 20742, USA*

*\*Corresponding author: cwu2011@umd.edu*

We have designed a plenoptic sensor to retrieve phase and amplitude changes resulting from a laser beam's propagation through atmospheric turbulence. Compared with the commonly restricted domain of ($-\pi$, $\pi$) in phase reconstruction by interferometers, the reconstructed phase obtained by the plenoptic sensors can be continuous up to a multiple of $2\pi$. When compared with conventional Shack-Hartmann sensors, ambiguities caused by interference or low intensity, such as branch points and branch cuts, are less likely to happen and can be adaptively avoided by our reconstruction algorithm. In the design of our plenoptic sensor, we modified the fundamental structure of a light field camera into a mini Keplerian telescope array by accurately cascading the back focal plane of its object lens with a microlens array's front focal plane and matching the numerical aperture of both components. Unlike light field cameras designed for incoherent imaging purposes, our plenoptic sensor operates on the complex amplitude of the incident beam and distributes it into a matrix of images that are simpler and less subject to interference than a global image of the beam. Then, with the proposed reconstruction algorithms, the plenoptic sensor is able to reconstruct the wavefront and a phase screen at an appropriate depth in the field that causes the equivalent distortion on the beam. The reconstructed results can be used to guide adaptive optics systems in directing beam propagation through atmospheric turbulence. In this paper we will show the theoretical analysis and experimental results obtained with the plenoptic sensor and its reconstruction algorithms.

## 1. Introduction

Atmospheric turbulence is a natural phenomenon that causes tiny inhomogeneity in the refractive index of air. It is frequently observed in our daily life such as when watching a star twinkle or observing an object shimmering on a hot day. Generally speaking, the magnitude of fluctuations in the atmospheric turbulence is less than $10^{-6}$, as tiny as one millionth of its normal value. However, these insignificant changes make air an effective shield against long range laser weapons[1,2] and simultaneously a significant channel disturbance in free space optic (FSO) communications[3,4].

Adaptive optics (AO) has been used to correct distorted wavefronts[5]. Normally a deformable mirror is used to compensate the distorted wavefront at the receiver end of the observing system. It requires guidance to operate, such as using a reference star to perform deconvolutions in astronomy[6], using a Shack-Hartmann sensor[7] to measure local tilts of the wavefront, or following certain deformation sequences to distinguish signal from noise[8]. Adaptive optics works efficiently in weak turbulence situations[9]. However, in stronger and deeper turbulence, phenomena such as beam break-up, scintillations[10] and self-interference[11] are commonly observed and can't be recognized accurately by a Shack-Hartmann sensor. Also, the continuous surface of a deformable mirror can't correct discontinuous wavefronts.

An advanced approach is to predistort the beam[12] with superposed time varying phase information that will be compensated by the turbulence during propagation, and then a relatively undistorted beam will arrive at the receiver. This approach is inspired by the intrinsic reciprocity[13] of atmospheric turbulence, and requires accurate measurements of the beam's phase and amplitude distributions. Therefore, a sensor that can detect and reconstruct complex amplitude of the field[14] is essential to the success of this approach.

We developed our plenoptic sensor and its associated algorithms[15] to achieve observation and extraction of the characteristics of a coherent light field. In principle, the hardware of our plenoptic sensor has evolved from the light field camera developed for computer vision applications[16,17]. The optic components are used as hardware layer computation tools while the algorithms of our plenoptic sensor are related to those used with Shack-Hartmann sensors but with swapped angular and spatial resolutions[18]. The mechanisms, algorithms and experimental results obtained with our plenoptic sensor will be presented in Part 2, 3, and 4, respectively.

## 2. Structure and Mechanisms of our Plenoptic Sensor

The basic idea in our design of the plenoptic sensor is to use all the lens components as Fourier transform tools that operate on the complex amplitude of the light field. A fundamental conclusion in Fourier optics states that in the paraxial approximation, the complex amplitude of light fields at the front and back focal planes of a lens are Fourier transforms of each other[19,20]. The terms "front" and "back" are dictated by the propagation direction of the light.

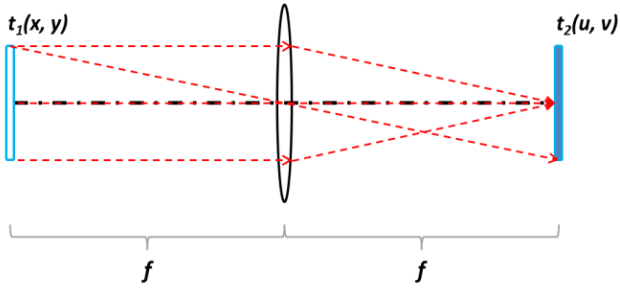

Figure 1: Fourier transform pairs of complex amplitude in the light fields at focal planes of a thin lens (the ray tracing shows the transforms between a point source and a planewave)

An analytical formula for the Fourier transform of a thin lens is expressed as[19]:

$$t_2(u,v) = \frac{1}{j\lambda f} \int_{-\infty}^{+\infty}\int_{-\infty}^{+\infty} t_1(x,y) \exp\left(-j\frac{2\pi}{\lambda f}(xu+yv)\right) dx dy \quad (1)$$

In equation (1), $t_1(x, y)$ and $t_2(u, v)$ are the complex amplitude of the field at the front and back focal planes of the thin lens, respectively. The focal length is represented by $f$. A Fourier transform is achieved by regarding the spatial frequency components as[19]:

$$f_x = \frac{u}{\lambda f}, \quad f_y = \frac{v}{\lambda f} \quad (2)$$

Thus, neglecting aperture limiting effects, the Fourier transform conducted by a thin lens swaps the geometric and angular information of the incident light field. The structure diagram of our plenoptic sensor is shown in Figure 2 as[18]:

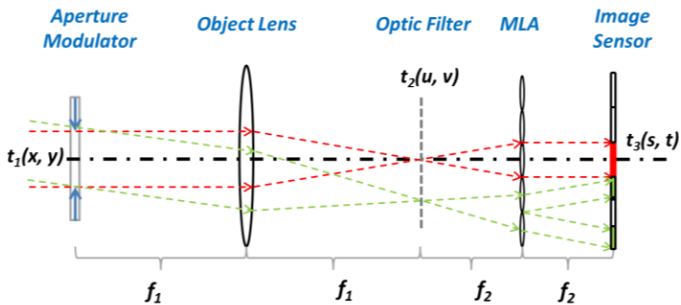

Figure 2: Structure Diagram of Plenoptic Sensor

In Figure 2, the front focal plane of the microlens array (MLA) coincides with the back focal plane of the objective lens. Their numerical apertures should satisfy:

$$NA_{objective} \leq NA_{MLA} \quad (3)$$

The numerical aperture of the MLA is defined by treating each microlens in the MLA (MLA cell) as individual lenses. In order to achieve the largest field of view, we make the numerical apertures of the MLA and the objective lens equal. If an arbitrary thin lens is used as the objective lens, the aperture modulator can change the effective numerical aperture of the objective lens to satisfy equation (3). The optic filter in figure 2 is a protection layer against strong intensities. For low power lasers, the optic filter can be omitted. Intuitively, the structure of the plenoptic camera looks like an array of Keplerian telescopes[21] that shares a common objective lens.

The wave analysis of the objective lens in the plenoptic sensor is the same as equation (1). However, the Fourier transform by each MLA cell should consider the aperture limiting effects since the width of the light field at the back focal plane of the objective lens is larger than the width of a MLA cell. Thus, a pupil function should be added to the integral when applying equation (1). Without loss of generality, one can express the field of $t_3(s, t)$ as a superposition of transforms performed by each MLA cell:

$$t_3(s,t) = \sum_{M,N} t_3^{M,N}(s',t') \quad (4)$$

In equation (4), the integer pair $(M, N)$ corresponds to the index for each MLA cell in a Cartesian layout. $(s', t')$ are the local coordinates in the domain of each MLA cell with relation to the "global" coordinates $(s, t)$, as:

$$(s',t') = (s - Md, t - Nd) \quad (5)$$

Symbol $d$ in equation (5) represents the pitch of the MLA (spacing between neighboring micro-lens centers). Therefore, the field at the back focal plane of the MLA for each micro-lens can be solved as[18]:

$$t_3^{M,N}(s',t') = \frac{1}{j\lambda f_2} \iint t_2(u'+Md, v'+Nd) \, rect\left(\frac{u'+s'}{d}\right) \\ rect\left(\frac{v'+t'}{d}\right) e^{-\frac{j2\pi}{\lambda f_2}(u's'+v't')} du' dv' \quad (6)$$

In equation (6), $rect(*)$ denotes the rectangular pupil function for each MLA cell. $(u', v')$ are local coordinates for the light field $t_2(u, v)$ viewed by each MLA cell that satisfy the relation:

$$(u',v') = (u - Md, v - Nd) \quad (7)$$

The aperture limiting effect is generally regarded in optics as the "vignetting" effect to indicate the reduced effective aperture for off-axis points. It is reflected in equation (6) by the pupil function. Often "vignetting" effects are treated as disadvantages that should be carefully avoided or compensated for in optic designs. However, these effects serve as an inter-block relation in the plenoptic sensor since the coordinates $(u', v')$ and $(s', t')$ are included in the same pupil function. In other words, each point $(s', t')$ in a MLA cell collects information from a slightly different area on the field of $t_2(u, v)$, where $(u, v)$ represent the angular spectrum of the field. Thus the "vignetting" effects provide finer adjustments of angular information in addition to the information provided by the index of the MLA cell $(M, N)$.

Combining equation (4) and (6) one can derive the wave solution for the plenoptic sensor. Due to the limited range of $(M, N)$, one can swap the order of summation and integration. As a result, the general solution is expressed as:

$$t_3(s,t) = \frac{1}{j\lambda f_2} \iint t_2(u,v) \sum_{M,N} rect\left(\frac{u+s-2Md}{d}\right) rect\left(\frac{v+t-2Nd}{d}\right) \quad (8)$$
$$\exp\left\{-j\frac{2\pi}{\lambda f_2}[(u-Md)(s-Md)+(v-Nd)(t-Nd)]\right\} dudv$$

An intuitive observation from equation (8) is that the effective integral area is a square of size $d \times d$ ($d$ is the pitch length of the MLA) for any pixel with fixed coordinates $(s, t)$. Each integral is based on the value of $t_2(u, v)$ with a linear phase tilt modulation. Thus the intensity obtained on $I_3(s, t)$ is the magnitude of a local Fourier transform with a linear geometric shift depending on the value of $(s, t)$.

For example, if the incoming light field consists of a group of interfering light patches (small "plane waves" with apertures), $t_2(u, v)$ will be a sum of delta functions in the form:

$$t_2(u,v) = \sum_{i=1}^{N} A_i e^{j\varphi_i} \cdot \delta(u-u_i, v-v_i) \quad (9)$$

After propagation through the MLA, the situation can be classified into 2 major cases:

Case 1: All the $(u_i, v_i)$ are distinctive enough that they are observed by different MLA cells.

Case 2: There exists more than one pair $(u_i, v_i)$ that falls in the same domain of a single MLA cell.

In case 1, one can easily determine the first order tilts in the complex amplitude of the field as each patch is imaged by an individual MLA cell. Thus the complex amplitude can be expressed as:

$$t_1(x,y) = \sum_{i=1}^{N} \gamma_0 \frac{\sqrt{I_i} f_2}{f_1} \exp\left(j2\pi d \frac{M_i x + N_i y}{\lambda f_1} + j\varphi_i\right) \quad (10)$$

In equation (10) $\gamma_0$ is a constant coefficient relating the optic field strength to the square root of pixel values. $(\gamma_0)^2$ represents the ratio between local wave intensity and corresponding pixel value on an image sensor. $I_i$ is the pixel value for the $i^{th}$ patch that represents the intensity. We arbitrarily neglect the intensity distribution to emphasize the capability of the plenoptic sensor in extracting the phase gradient. In fact, the intensity distribution is preserved under the transforms of the plenoptic sensor's lens system down to the limit of pixel sizes. However the initial phase information (DC value of phase) is lost as the patches don't interfere with each other when imaged by different MLA cells.

In case 2, if more than one patch propagates through the same MLA cell, their initial phase difference as well as their first order phase tilts can be revealed. Without loss of generality, assume 2 patches with amplitude $A_1$ and $A_2$ and phase difference $\Delta\varphi$ are observed by the same MLA cell. Then, the complex amplitude after the MLA cell can be expressed as:

$$t_3^{M,N}(s',t') = \frac{A_1}{j\lambda f_2} rect\left(\frac{u_1+s'-Md}{d}\right) rect\left(\frac{v_1+t'-Nd}{d}\right)$$
$$+ \frac{A_2 e^{j\Delta\varphi}}{j\lambda f_2} rect\left(\frac{u_2+s'-Md}{d}\right) rect\left(\frac{v_2+t'-Nd}{d}\right) \cdot \quad (11)$$
$$\exp\left\{j\frac{2\pi}{\lambda f_2}[(u_2-u_1)s'+(v_2-v_1)t']\right\}$$

Note that we have ignored the common phase that has no influence on the image. The corresponding sub-image can be written as:

$$I_3^{M,N}(s',t') =$$
$$\begin{cases} \eta A_1^2 & (u_1',v_1') \in \{A(s',t')\} \text{ while } (u_2',v_2') \notin \{A(s',t')\} \\ \eta A_2^2 & (u_1',v_1') \notin \{A(s',t')\} \text{ while } (u_2',v_2') \in \{A(s',t')\} \\ \eta[A_1^2+A_2^2+2A_1 A_2 \cos(\Delta u \cdot s'+\Delta v \cdot t'+\Delta\varphi)] & \text{both} \in \{A(s',t')\} \end{cases} \quad (12)$$

In equation (12), $\{A(s', t')\}$ is the integral area determined by $(s', t')$ from equation (6). We use $\eta$ as a coefficient representing the linear relation between pixel value and field intensity to simplify the result. Thus if the 2 patches are imaged by the same MLA cell, their initial phase difference as well as their first order phase tilt can be retrieved.

An overall relation between complex amplitude of the field and the final image can be derived by combining equation (1) and equation (8). The final result is expressed as:

$$t_3(s,t) = -\frac{1}{\lambda^2 f_1 f_2} \sum_{M,N} \int_{2Md-d/2-s}^{2Md+d/2-s} \int_{2Nd-d/2-t}^{2Nd+d/2-t} \int_{-\infty}^{+\infty} \int_{-\infty}^{+\infty} t_1(x,y)$$
$$\exp\left(-j\frac{2\pi}{\lambda f_1}(xu+yv)\right) \cdot \exp\left(-j\frac{2\pi}{\lambda f_2}(u-Md)(s-Md)\right) \cdot \quad (13)$$
$$\exp\left(-j\frac{2\pi}{\lambda f_2}(v-Nd)(t-Nd)\right) dxdydudv$$

Simpler results can be derived using geometrical optics expressed in the form of a plenoptic function[21] and neglecting the small apertures of the MLA cells:

$$\rho\left(s,t,\frac{ds}{dz},\frac{dt}{dz},\phi; z=2f_1+2f_2\right) = \frac{f_1}{f_2} \rho\left(x,y,\frac{dx}{dz},\frac{dy}{dz},\phi_0; z=0\right) \quad (14)$$

In equation (14) $\rho$ is the square root intensity of an arbitrary ray in the light field that enters the plenoptic sensor. $(s, t)$ and $(ds/dz, dt/dz)$ are geometric and angular information for the ray respectively on the transverse plane at depth $z$. $\phi$ is the phase of the ray. Its parameters satisfy the following relations[18]:

$$(s,t) = d \cdot (M,N) - \frac{f_2}{f_1}(x,y) \quad (15)$$

$$\left(\frac{ds}{dz},\frac{dt}{dz}\right) = \frac{nd}{f_2}(M,N) - \frac{f_1}{f_2}\left(\frac{dx}{dz},\frac{dy}{dz}\right) \quad (16)$$

$$(M,N) = \left\lfloor \left(\frac{x}{d}+\frac{1}{2},\frac{y}{d}+\frac{1}{2}\right) \right\rfloor \quad (17)$$

$$\frac{\partial[\phi(s,t)-\phi_0(x,y)]}{\partial(dx/dz)} = \frac{\partial[\phi(s,t)-\phi_0(x,y)]}{\partial(dy/dz)} = 0 \quad (18)$$

It is not surprising to find that the geometrical optic approach provides results very close to the analytical result from the wave solution approach, unless the wavefront oscillates at a high spatial frequency. Thus either approach can be used for extracting light field information from the image obtained by the plenoptic sensor.

## 3. Reconstruction Algorithms

The plenoptic sensor operates like a light field analyzer. In fact, it has the following functions:
(1) The objective lens sorts the angular information of the incident light field (or beam) into different

corresponding block images (sub-images) marked with index pair *(M, N)*.
(2) The MLA retrieves the spatial information of the incident light field by a linear scaling factor *(-f_2/f_1)*.
(3) The relative phase information is preserved for rays entering the same spot on the front focal plane of objective lens and ending in the same pixel on the image sensor.

Equivalently, one can explain the work of the plenoptic senor as translating the resolution of a 2D image into a 4D phase space of the light field (2D for geometric locations and 2D for angular information). The block index *(M, N)* are the indicators of each ray's dominant angular information while the local coordinates *(s', t')* in the domain of each MLA cell indicate the geometric information. Therefore, a complex, coherently interfering, light field is separated into an array of block images. There are fewer interfering patches in each block image compared to an image from a normal camera. Information from the patches can be retrieved from the block images as demonstrated in part 2. Thus the entire light field can be reconstructed with reasonable accuracy.

Several reconstruction algorithms are developed for various applications:

### Algorithm (1): Single phase screen reconstruction

If a small segment along the propagation distance is known to have significantly higher turbulence, it can be modeled as a phase screen. The plenoptic sensor can be used to reconstruct the phase screen at the dominant turbulence location. The compensation strategy is to arrange the deformable mirror's (DM) surface so that the imposed phase distortion on the outgoing beam will be largely cancelled out by the propagation through turbulence.

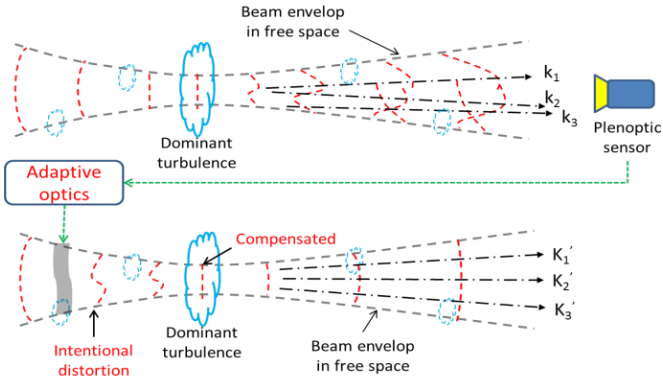

Figure 3: Phase compensation algorithm diagram for beam distortion caused by dominant turbulence occurring over a short distance

Figure 3 demonstrates how to use a plenoptic camera to determine the phase distortion of a beam propagated through atmospheric turbulence (upper part of the figure). The reconstructed phase information is sent to the AO device to compensate for the turbulence effect[22]. Thus a less distorted beam can be obtained at the receiver site after the one-step correction (lower part of the figure).

The corresponding image processing algorithm uses the following steps:
1. Select an MLA cell and its corresponding block image as a geometric reference.
2. Shift all nonzero block images to the block of the reference image and extend the scalar pixel values to a cluster of vectors with baseline directions extracted from MLA index *(M, N)* and their length proportional to the pixel brightness (ray intensity).
3. Adjust the vectors' directions according to their relative locations in the block due to the "vignetting" effect, or alternatively use equation (16).
4. Back propagate the rays to the depth of the optic event (dominant turbulence location).
5. Filter out rays that are traced out of the reasonable range and rays with abruptly different angles from their geometric neighboring rays (ignore unreasonable rays).
6. Project the beam propagation to the reconstruction plane assuming no turbulence and reshape it in a vector form with the same geometric resolution as in step 4.
7. Combine the ray patterns before and after the "phase screen" to extract the gradients of the phase screen.
8. Build the phase screen according to its gradient profile.

Intuitively, the gradients of the phase will cause variations of Poynting vectors in the wave that can be picked up by the plenoptic sensor. Therefore, the phase screen's scattering patterns can be largely retrieved by backward ray tracing. The gradient of phase distortion can be reconstructed as:

$$\nabla_{i,j}\phi(x, y; z = z') = -\frac{1}{2}\nabla_{i,j}\phi_{project}(x, y; z = z')$$
$$+ \frac{\iint_{I>I_{th}} I(s,t; z = 2f_1 + 2f_2)\left[\frac{2nd \cdot (M,N) - n \cdot (s,t)}{f_1}\right]dsdt}{2\iint_{I>I_{th}} I(s,t; z = 2f_1 + 2f_2)dsdt} \quad (19)$$

The integral area in equation (19) is the area of the PSF (point spread function) for each point source located at *(x, y; z=z')* with z=z' indicating the plane of reconstruction. In addition, due to the continuity of a phase screen, an extra layer of filtering can be applied based on the fact that any integral loop of the phase gradient equals zero. In fact, this law should be satisfied for all reconstruction algorithms of a continuous phase screen.

### Algorithm (2): Intelligent interferometer

The plenoptic sensor can be used as an intelligent interferometer that doesn't require a reference beam. The reconstruction process can determine the phase changes as well as the amplitude changes of a distorted beam. Therefore, a surface structure with tiny variations (on the order of 1μm) can be illuminated by a coherent beam and reconstructed with the plenoptic sensor instead of an interferometer. To achieve maximum accuracies, the detected surface should be placed at the front focal plane of the objective lens.

The algorithm steps are as follows:
1. Scale the geometry of all nonzero block images by *(-f_1/f_2)*.
2. Retrieve rays' angular information based on equation *(15)*, *(16)* and *(17)*.
3. Apply the result from step 2 to equation *(19)* to determine the gradients of surface.
4. Reconstruct the optic surface based on the surface's gradient profile.

It is not difficult to see that this algorithm is a simplified version of algorithm (1) without the back propagation process. A detailed demonstration and discussion of this algorithm and its advantages compared with conventional interferometers are included in part 4 based on a laboratory generated deformation.

### Algorithm (3): Multi phase screens reconstruction

Algorithm (1) requires that the major turbulence effects are caused by a small segment of the channel. Algorithm (2) requires placing the surface to be reconstructed at the front focal plane of the plenoptic sensor. However, in more general cases, there is no knowledge or restriction on the turbulence channel when a distorted beam is observed by the plenoptic sensor. Can the plenoptic sensor still provide some valuable information for the adaptive optics system?

The short answer is YES. In fact, 2 scenarios can be considered. On one hand, we can treat the atmospheric turbulence effect as an outcome of one phase screen (2D model) at an arbitrary depth in the channel. This scenario is covered by Algorithm (1): single phase screen reconstruction. On the other hand, we can treat the distortion as the combined effects of several distributed phase screens (3D model). This second scenario is aimed at using one plenoptic image to reconstruct several phase screens. The challenge in the second scenario is to distinguish the parts on a plenoptic image that help to retrieve a distributed phase screen from the parts that won't help.

This challenge can be illustrated by figure 4:

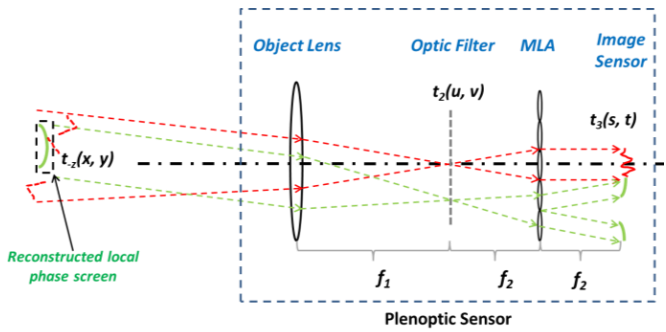

Figure 4: Diagram for reconstructing distributed simple and small phase screens in 3D

In figure 4, the green parts help to reconstruct one of the distributed phase screens while the red parts don't. Therefore, the reconstruction of this phase screen only requires part of the plenoptic image.

The algorithm steps are:
1. Select a group of block images that contain large uniform (less interfered) light patches and mark them as set *{U}*.
2. Mark the remaining nonzero block images as set *{V}*.
3. Back trace all the light patches in image set *{U}* by a finite step distance and calculate the spatial variance of intensity sums.
4. If the variance of the intensity sums at this new location is lower than that of its block image components, it implies a more uniform image can be formed. Therefore, a small and simple phase screen can be reconstructed at this new location by applying algorithm (1). The light patches back traced through this phase screen should be adjusted accordingly. Remove these patches from set *{U}*.
5. Go back to step 3 and continue the back tracing process until the transmitter's site is reached or the number of elements in *{U}* drops below two (no more light patches to combine).
6. Back propagate the small light patches in set *{V}* to examine the correctness of reconstruction. Delete the phase screens that don't intercept any of the small light patches in set *{V}*.

## 4. Experiment results

In experiments, we used an OKO deformable mirror[23] (piezoelectric, 37 actuators, 30mm diameter) to generate known phase distortions. We then use the plenoptic sensor to reconstruct the phase distortion independently. A collimated Gaussian beam is used as the transmitted beam. The experiment diagram and layout are as shown:

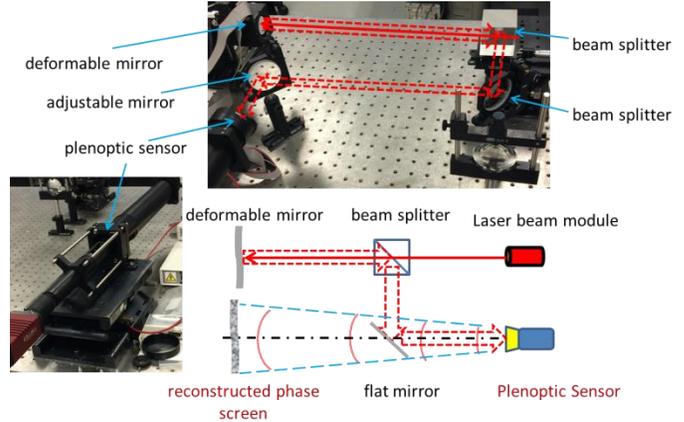

Figure 5: Experiment diagram and layout for the plenoptic sensor

The plenoptic sensor contains a thin object lens with a 2 inch diameter and focal length of 750mm. The MLA used is 10mm by 10mm with 300μm pitch length and 1.6° divergence (EF = 5.1mm). The image sensor is an Allied Vision GX1050 monochrome CCD camera with a max frame rate at 109fps. A detailed diagram of the OKO 37-channel deformable mirror and its actuators' locations are shown in Figure 6.

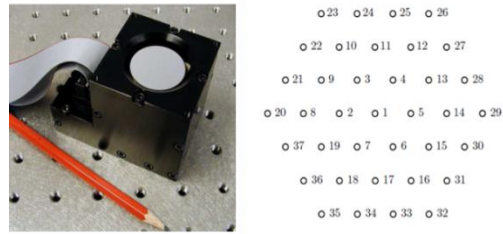

Figure 6: OKO 37-channel PDM and its actuators' positions (observed from the back of the mirror)

The commands sent to the deformable mirror consist of 37 individual integer numbers ranging from 0 to 4095. The maximum number 4095 corresponds to a voltage of 150V, which drives a piezoelectric actuator in the DM to move to its full magnitude of 5.5λ (λ=632.8nm). When a deformation needs to be applied by the 37-actuators beneath the surface of the DM, one can use the experimentally determined scaling factor of 740/λ to form the distortion command.

The results for the reconstruction algorithms are shown as:
### 1. Algorithm(1): Single phase screen reconstruction algorithm:

Case A:     Defocus Z(2, 0) deformation

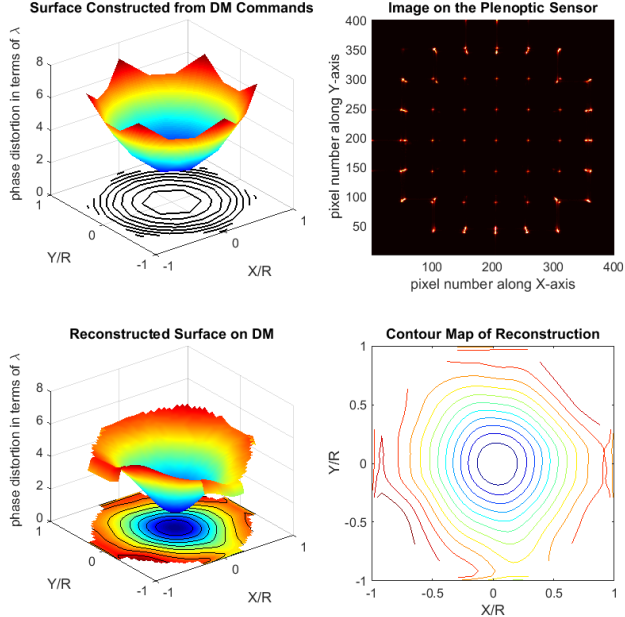

Figure 7: Reconstruction results of a phase screen with Z(2, 0)

In figure 7, the upper-left plot shows the shape and magnitude of the phase distortion that is applied to the deformable mirror. A contour plot of the deformation is shown on *X-Y* plane.

The upper-right plot in figure 7 shows the image on the plenoptic sensor when a "Defocus" command is sent to the DM. We only show the illuminated blocks on the image sensor. The size of the image sensor (resolution=1024×1024, pixel pitch=5.5μm) supports a maximum number of 18×18 blocks of sub-images. Equivalently, the maximum detectable distortion for the plenoptic sensor is ±1.4λ/mm. In the case of the "Defocus", the furthest block from the center is *(M=4, N=0)* and the corresponding maximum tilt can be calculated as 0.6λ/mm. The "Defocus" can be expressed as:

$$Z_2^0(\rho, \theta) = A \cdot \rho^2 \qquad 0 \leq \rho \leq 1, \ \theta \in [0, 2\pi) \qquad (20)$$

The symbol *"A"* in equation (20) represents the magnitude of the distortion (*A=4095* in the case of "Defocus"), while $\rho$ and $\theta$ represent the normalized radius and angle respectively for the Zernike polynomial. Intuitively, the gradient of the "Defocus" function increases symmetrically when the radius $\rho$ increases. The gradient at each spot $(\rho, \theta)$ is mapped into different blocks of the deformable mirror. Furthermore, the observation that the most outside blocks are illuminated with larger areas reflects that the gradients changes faster when the radius $\rho$ increases.

The lower-left plot in figure 7 shows the reconstruction result of the deformable mirror's surface. According to the algorithm steps of single phase screen reconstruction (algorithm 1), the center block of the plenoptic image is selected as the reference block and the reconstruction achieved by examining all the illuminated blocks. The clipping at the edges of the reconstruction are because of the lack of boundary conditions. There is no information about further phase variation outside the edge of the reconstructed surfaces. Therefore, we simply set the reconstructed phase value outside the edges to be zero. A contour plot is presented on the *X-Y* plane.

The lower-right plot in figure 7 is a detailed contour map of the reconstructed surface. The contour plot projects the 3D result of the reconstructed phase screen into a 2D plot and helps to show the details of the reconstruction. In the case of "Defocus", the contour plot shows the concentric rings of the deformation. It looks similar to the contour plot of the commands at the upper-left plot in figure 7.

Using the reconstruction algorithm, one can determine:
(1) The geometrical information of light patches (rays) at the plane of distortion. (Algorithm steps 1, 2 and 4)
(2) The angular information of light patches (rays) at the plane of distortion (Algorithm steps 1, 2 and 3)
(3) The phase gradient along the X axis and Y axis. (Algorithm steps 5, 6 and 7)

A detailed explanation of the algorithm can be illustrated by figure 8:

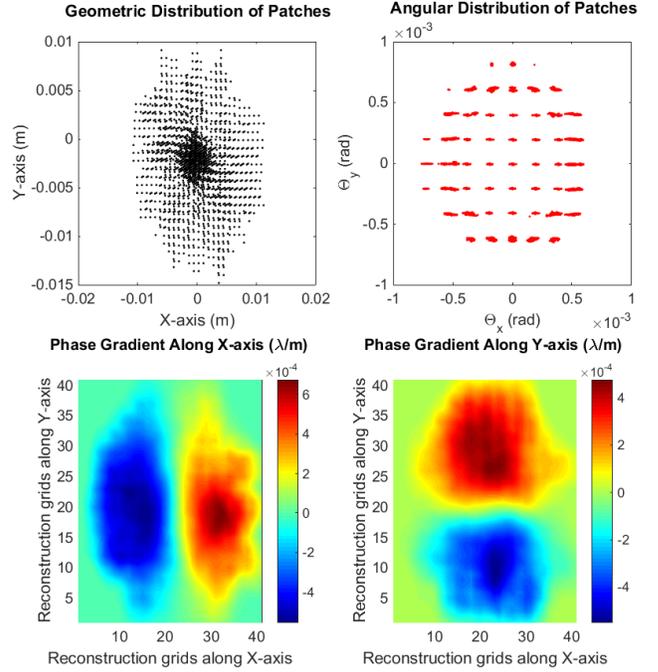

Figure 8: Reconstruction details for case "Defocus"

In figure 8, the upper-left plot shows the geometric distribution of light patches at optimized back tracing depth, with each dot representing a small patch (or single ray). The optimized depth (the plane of reconstruction) is determined by back propagating the rays until the intensity distribution resembles the beam profile before encountering the phase screen (a Gaussian distribution in our experiment). The upper-right plot shows the distribution of the directions of the patches (rays), with each dot representing a small patch (or single ray). The angular distribution of patches results from the gradient of the phase change and is extracted from the image on the plenoptic sensor primarily by the block index *(M, N)* and adjusted by the actual positions of rays on the reconstruction plane. Based on the geometric and directional information of light patches (rays), the phase gradient graphs can be determined by equation (19). The results of the phase gradient along the X axis and Y axis are presented in the lower-left and lower-right plots in figure 8 respectively. With all the necessary information, algorithm step 8 can be completed to derive the results demonstrated in figure 7.

Based on the reconstruction results, the channel (actuator) values on the deformable mirror can be quantitatively extracted by sampling the reconstructed surface at the geometries of the 37-channel actuators of the DM. Then by multiplying the same scaling factor of 740/λ, the channel based values can be compared with the original commands sent to the deformable mirror. The compact results in the "Defocus" case is listed in the table 1:

Table 1. Channel values of "Defocus"

| Channel | Distortion | Reconstruction | Error |
|---|---|---|---|
| 1 | 0 | 0 | 0 |
| 2 | 455 | 742 | 287 |
| 3 | 455 | 641 | 186 |
| 4 | 455 | 633 | 178 |
| 5 | 455 | 782 | 327 |
| 6 | 455 | 916 | 461 |
| 7 | 455 | 794 | 339 |
| 8 | 1365 | 1828 | 463 |
| 9 | 1820 | 2276 | 456 |
| 10 | 1365 | 1862 | 497 |
| 11 | 1820 | 2267 | 447 |
| 12 | 1365 | 1904 | 539 |
| 13 | 1820 | 2225 | 405 |
| 14 | 1365 | 1716 | 351 |
| 15 | 1820 | 2247 | 427 |
| 16 | 1365 | 2064 | 699 |
| 17 | 1820 | 2611 | 791 |
| 18 | 1365 | 2045 | 680 |
| 19 | 1820 | 2149 | 329 |
| 20 | 3185 | 2671 | -514 |
| 21 | 3185 | 2977 | -208 |
| 22 | 4095 | 3443 | -652 |
| 23 | 3185 | 3005 | -180 |
| 24 | 3185 | 2976 | -209 |
| 25 | 4095 | 3730 | -365 |
| 26 | 3185 | 3302 | 117 |
| 27 | 3185 | 3229 | 44 |
| 28 | 4095 | 3453 | -642 |
| 29 | 3185 | 2963 | -222 |
| 30 | 3185 | 2821 | -364 |
| 31 | 4095 | 3361 | -734 |
| 32 | 3185 | 3167 | -18 |
| 33 | 3185 | 3313 | 128 |
| 34 | 4095 | 4095 | 0 |
| 35 | 3185 | 3374 | 189 |
| 36 | 3185 | 3238 | 53 |
| 37 | 4095 | 3040 | -1055 |

The first column of table 1 is the index of the actuators as demonstrated as in figure 8. The second column is the commands sent to the actuators of the DM, and the third column is the information extracted from the reconstruction at each position of the actuators. The fourth column (Error) in table 1 is derived by subtracting the third column by the second column. It reflects the overshoot of the reconstruction result compared with the original commands sent to the deformable mirror. Compared with the scaling factor of *740λ*, most channels are reconstructed within an absolute error of *λ/2* when the maximum deformation (*5.5λ*) is applied. Without loss of generality, we can use the correlation between the original commands and the reconstruction results to reflect how closely the deformation is recognized. A correlation value of *"1"* represents the complete recognition of the distortion; while a correlation value of *"0"* means that the reconstructed surface is completely irrelevant to the actual distortion. In the "Defocus" case, the correlation value is 0.956. According to figure 3, once the distortion pattern is recognized, the AO device can be intelligently arranged to compensate for the major phase distortion. The remaining distortion can be corrected by iteratively using the sensing, reconstruction, and compensation process, or by using conventional AO strategies such as SPGD[24] or SIMPLEX[25] methods.

The time consumption in a control-feedback loop in our experiment is about 13ms, which includes the image acquisition time (9.2ms), the algorithm processing time (3.1ms on CPU) and setup time for the DM (≈1ms). These times do not reflect the faster performance that could be achieved with more advanced (and costlier) components.

Other cases of Zernike polynomials expressed as *Z(m, n)* or equivalently $Z_n^m$ can be sensed and reconstructed accordingly. For example, figure 9-12 show the reconstruction results of *Z(1, 1), Z(2, 2), Z(3, 3)* and *Z(4, 4)* respectively:

Case B: Tilt *Z(1, 1)* deformation

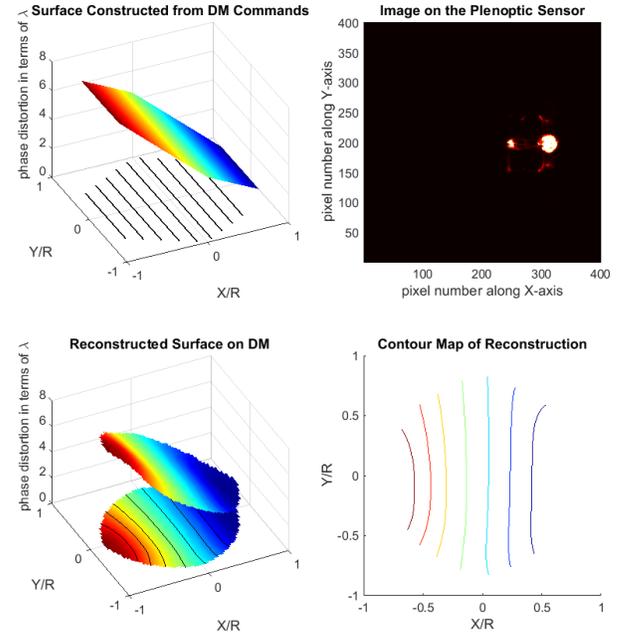

Figure 9: Reconstruction results of a phase screen *Z(1, 1)*

The function *Z(1, 1)* can be expressed as:

$$Z_1^1(\rho,\theta) = A \cdot \rho \qquad 0 \le \rho \le 1, \quad \theta \in [0, 2\pi) \qquad (21)$$

Case C: Astigmatism *Z(2, 2)* deformation

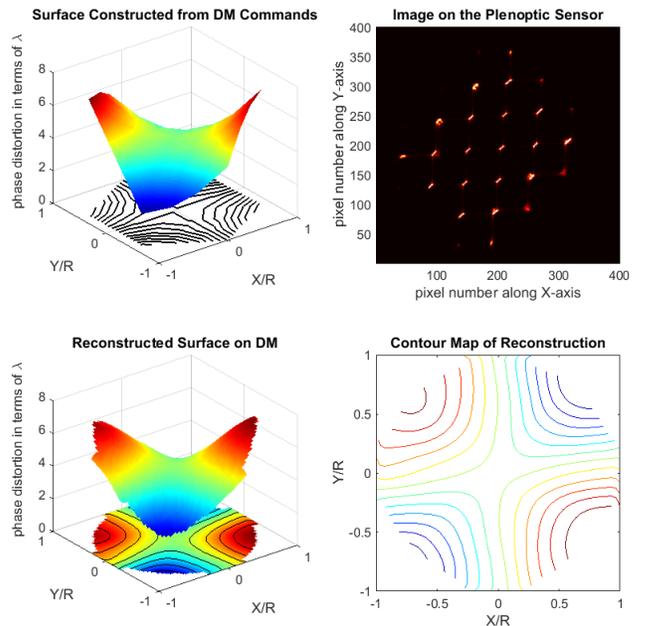

Figure 10: Reconstruction results of a phase screen *Z(2, 2)*

The function $Z(2, 2)$ can be expressed as:

$$Z_2^2(\rho,\theta) = \frac{A}{2}\left[\rho^2 \cdot \cos(2\theta) + 1\right] \quad 0 \le \rho \le 1, \ \theta \in [0, 2\pi) \quad (22)$$

Case D:        Trefoil $Z(3, 3)$ deformation

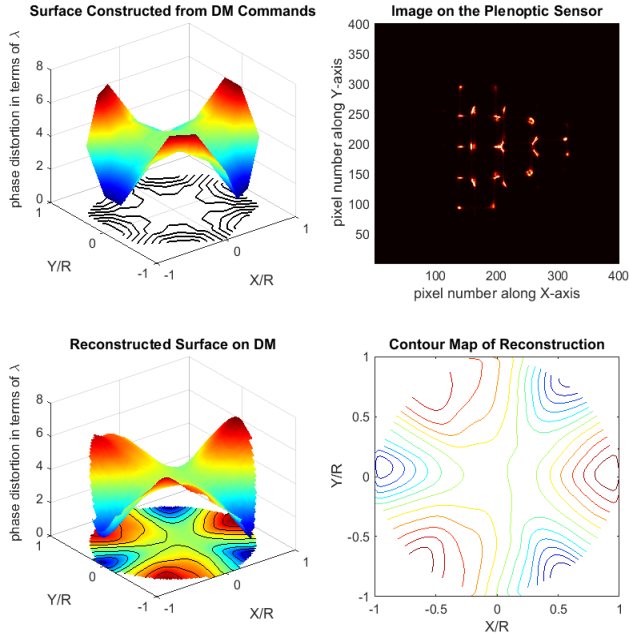

Figure 11: Reconstruction results of a phase screen $Z(3, 3)$

The function $Z(3, 3)$ can be expressed as:

$$Z_3^3(\rho,\theta) = \frac{A}{2}\left[\rho^3 \cdot \cos(3\theta) + 1\right] \quad 0 \le \rho \le 1, \ \theta \in [0, 2\pi) \quad (23)$$

Case E:        Tetrafoil $Z(4, 4)$ deformation

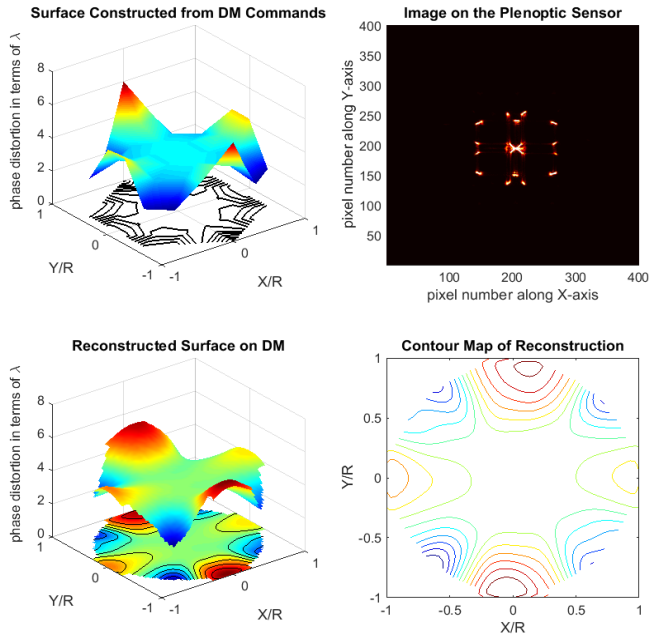

Figure 12: Reconstruction results of a phase screen $Z(4, 4)$

The function $Z(4, 4)$ can be expressed as:

$$Z_4^4(\rho,\theta) = \frac{A}{2}\left[\rho^4 \cdot \cos(4\theta) + 1\right] \quad 0 \le \rho \le 1, \ \theta \in [0, 2\pi) \quad (24)$$

The accuracies of the above reconstructions (figure 9-12) can be evaluated similarly to the "Defocus" case by extracting the values for individual actuators and compared them with the original commands. The results are listed in table 2-3:

Table 2. Channel values of "Tilt" and "Astigmatism"

| | Z(1, 1) | | Z(2, 2) | |
|---|---|---|---|---|
| Channel | Distortion | Reconstruction | Distortion | Reconstruction |
| 1 | 2048 | 2037 | 2048 | 1797 |
| 2 | 1367 | 1224 | 1820 | 1521 |
| 3 | 1367 | 1195 | 2275 | 2191 |
| 4 | 2048 | 1928 | 2048 | 1917 |
| 5 | 2728 | 2796 | 1820 | 1541 |
| 6 | 2728 | 2796 | 2275 | 2343 |
| 7 | 2048 | 2054 | 2048 | 1883 |
| 8 | 1367 | 1250 | 1365 | 1017 |
| 9 | 686 | 575 | 1137 | 934 |
| 10 | 686 | 491 | 2048 | 2011 |
| 11 | 686 | 497 | 2958 | 3010 |
| 12 | 1367 | 1213 | 2730 | 2782 |
| 13 | 2048 | 1952 | 2048 | 2023 |
| 14 | 2728 | 2777 | 1365 | 1048 |
| 15 | 3409 | 3456 | 1137 | 950 |
| 16 | 3409 | 3493 | 2048 | 2164 |
| 17 | 3409 | 3526 | 2958 | 3327 |
| 18 | 2728 | 2817 | 2730 | 2701 |
| 19 | 2048 | 2092 | 2048 | 1832 |
| 20 | 1336 | 1428 | 865 | 727 |
| 21 | 711 | 886 | 236 | 68 |
| 22 | 5 | 583 | 0 | 230 |
| 23 | 0 | 106 | 1418 | 1797 |
| 24 | 0 | 0 | 2677 | 2678 |
| 25 | 5 | 6 | 4095 | 3622 |
| 26 | 711 | 544 | 3859 | 3658 |
| 27 | 1336 | 1249 | 3230 | 2972 |
| 28 | 2048 | 1997 | 2048 | 1953 |
| 29 | 2759 | 2791 | 865 | 828 |
| 30 | 3384 | 3375 | 236 | 0 |
| 31 | 4090 | 3783 | 0 | 216 |
| 32 | 4095 | 4095 | 1418 | 1905 |
| 33 | 4095 | 4084 | 2677 | 3054 |
| 34 | 4090 | 4004 | 4095 | 4095 |
| 35 | 3384 | 3508 | 3859 | 3640 |
| 36 | 2759 | 2787 | 3230 | 2715 |
| 37 | 2048 | 2117 | 2048 | 1678 |

Table 3. Channel values of "Trefoil" and "Tetrafoil"

| | Z(3, 3) | | Z(4, 4) | |
|---|---|---|---|---|
| Channel | Distortion | Reconstruction | Distortion | Reconstruction |
| 1 | 2048 | 2052 | 1484 | 1591 |
| 2 | 2048 | 1871 | 1468 | 1583 |
| 3 | 2048 | 1854 | 1468 | 1545 |
| 4 | 2048 | 2281 | 1516 | 1529 |
| 5 | 2048 | 2103 | 1468 | 1634 |
| 6 | 2048 | 2174 | 1468 | 1677 |
| 7 | 2048 | 2314 | 1516 | 1590 |
| 8 | 1384 | 1258 | 1339 | 1554 |
| 9 | 2048 | 1926 | 1226 | 1286 |
| 10 | 2711 | 2845 | 1774 | 1803 |
| 11 | 2048 | 2066 | 1226 | 1515 |
| 12 | 1384 | 1557 | 1339 | 1683 |
| 13 | 2048 | 2664 | 2000 | 2606 |
| 14 | 2711 | 2955 | 1339 | 1566 |
| 15 | 2048 | 2275 | 1226 | 1462 |
| 16 | 1384 | 1416 | 1774 | 1989 |
| 17 | 2047 | 2794 | 1226 | 1445 |
| 18 | 2711 | 2992 | 1339 | 1500 |
| 19 | 2048 | 2434 | 2000 | 2354 |
| 20 | 0 | 638 | 1758 | 2407 |
| 21 | 0 | 0 | 0 | 409 |
| 22 | 2047 | 2063 | 178 | 481 |
| 23 | 4095 | 3579 | 2694 | 1955 |
| 24 | 4095 | 3844 | 2694 | 2118 |
| 25 | 2048 | 2258 | 178 | 947 |
| 26 | 0 | 256 | 0 | 1027 |
| 27 | 0 | 1070 | 1758 | 3100 |
| 28 | 2047 | 2878 | 4095 | 3834 |
| 29 | 4095 | 3518 | 1758 | 2297 |
| 30 | 4095 | 3957 | 0 | 481 |
| 31 | 2048 | 2287 | 178 | 800 |
| 32 | 0 | 364 | 2694 | 2212 |
| 33 | 0 | 1396 | 2694 | 2242 |
| 34 | 2047 | 3269 | 178 | 1007 |
| 35 | 4095 | 4095 | 0 | 0 |
| 36 | 4095 | 3343 | 1758 | 2649 |
| 37 | 2048 | 2042 | 4095 | 4095 |

In table 2 and table 3, the "Distortion" columns are the commanded deformation values for the DM to enforce different Zernike Polynomials. The "Reconstruction" columns are the value of the actuators extracted from the reconstructed phase distortion. The reconstructed channel values closely resemble the imposed distortions.

The largest mismatch in the case "Tilt" is actuator #22 with absolute error of 578 (equivalent to 0.78λ). The largest mismatch in the case "Astigmatism" is actuator #36 with absolute error of 515 (0.7λ). For the case "Trefoil", the largest mismatch is actuators #33 with absolute error of 1396 (1.9λ). And for the case "Tetrafoil", the largest mismatch is actuator #27 with absolute error of 1342 (1.8λ). Although the deformable mirror's surface condition is invariant under a constant shift (add the same value to each actuator), the worst reconstructed values can serve as an upper limit of the reconstruction errors. All of the largest mismatched actuators are at the edge of the DM (actuator #20 to actuator #37). This can be explained by the fact that the reconstruction algorithm integrates the gradients of the phase distortion from the center point to the edge points of the DM's surface, and therefore the reconstruction errors accumulate and propagate to the outside actuators.

Table 4 provides the correlation and absolute RMS errors between the distortion values and reconstructed values for all achievable modes of Zernike polynomials on the deformable mirror. The mode $Z_4^0$ and modes higher than $Z_4^4$ can't be accurately enforced by the DM used in the experiment because the number of actuators in the DM is not adequate enough to represent spatial oscillations in the radius ($\rho$) direction.

Table 4. Difference between imposed distortions and reconstructed distortions in basic Zernike polynomials

| Zernike Polynomial | Correlation | RMS Error |
|---|---|---|
| Tilt ($Z_1^1$) | 0.993 | 0.20λ |
| Defocus($Z_2^0$) | 0.956 | 0.59λ |
| Astigmatism($Z_2^2$) | 0.975 | 0.32λ |
| Coma($Z_3^1$) | 0.936 | 0.49λ |
| Trefoil($Z_3^3$) | 0.932 | 0.62λ |
| Secondary Astigmatism ($Z_4^2$) | 0.842 | 0.60λ |
| Tetrafoil ($Z_4^4$) | 0.904 | 0.59λ |

In table 4, the "Correlation" column shows that how closely the mode is recognized compared to the reconstruction results. The results show that all the basic polynomial modes are recognized by the plenoptic sensor by the reconstruction algorithm. The "RMS Error" column demonstrates the average error for each reconstructed case of Zernike polynomials. Compared with the maximum phase distortion in each case (5.5λ), "RMS Error" accounts for on average 10% of the maximum magnitude of phase distortion.

To demonstrate a more general case where the distortion is a superposition of various Zernike polynomials, a case of half "Trefoil" + half "Tetrafoil" is shown as:

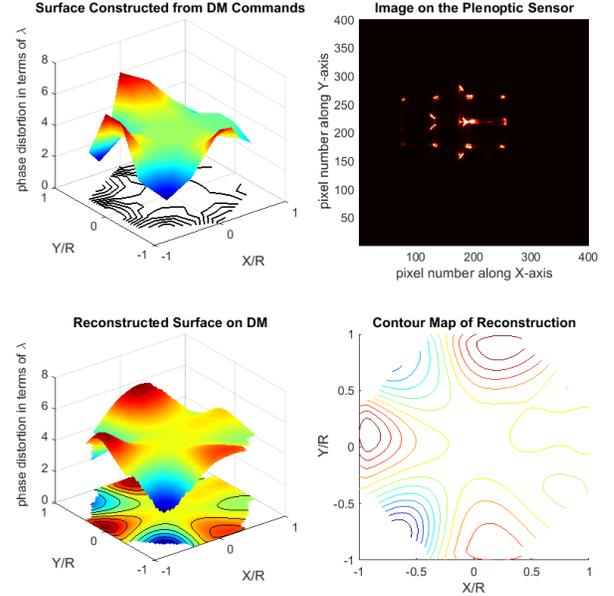

Figure 13: Reconstruction results of a phase screen
[Z(3, 3) + Z(4, 4)]/2

The function of the distortion used is:

$$Z(\rho,\theta) = \frac{A}{4}\left[ \rho^4 \cos(4\theta) + \rho^3 \cos(3\theta) + 2 \right] \quad (25)$$

where the parameters satisfy:

$$0 \le \rho \le 1, \ \theta \in [0, 2\pi) \quad (26)$$

Figure 13 shows that for an arbitrary distortion, the plenoptic sensor is able to use a single reconstruction process to recognize the major distortions in phase. In fact, the Zernike polynomials can be treated as Eigen-modes for a phase distortion and an arbitrary distortion can be expressed with combinations of basic Zernike polynomials plus small local distortions. Therefore, the reconstruction case in figure 13 is as accurate as the cases of single mode Zernike polynomials. The correlation between the initial distortion and the reconstructed results in terms of channel values is 0.917 and the "RMS Error" is 0.54λ.

The detailed channel values of the initial distortion and reconstructed distortion for figure 13 are listed in table 5:

Table 5. Channel values for combined Zernike Polynomial "Trefoil" + "Tetrafoil"

| Channel | Distortion | Reconstruction | Error |
|---|---|---|---|
| 1 | 2144 | 2266 | 122 |
| 2 | 2133 | 2232 | 99 |
| 3 | 2133 | 2330 | 197 |
| 4 | 2167 | 2237 | 70 |
| 5 | 2133 | 2317 | 184 |
| 6 | 2133 | 2239 | 106 |
| 7 | 2167 | 2400 | 233 |
| 8 | 2397 | 2551 | 154 |
| 9 | 1962 | 2016 | 54 |
| 10 | 1994 | 2375 | 381 |
| 11 | 1962 | 2428 | 466 |
| 12 | 2397 | 2419 | 22 |
| 13 | 2509 | 3005 | 496 |
| 14 | 1687 | 1864 | 177 |
| 15 | 1962 | 2196 | 234 |
| 16 | 2705 | 2724 | 19 |
| 17 | 1962 | 1696 | -266 |
| 18 | 1687 | 1911 | 224 |
| 19 | 2509 | 3023 | 514 |
| 20 | 3434 | 3283 | -151 |
| 21 | 2192 | 2419 | 227 |
| 22 | 1222 | 1842 | 620 |
| 23 | 1903 | 2126 | 223 |
| 24 | 1903 | 2488 | 585 |
| 25 | 1222 | 2107 | 885 |
| 26 | 2192 | 2701 | 509 |
| 27 | 3434 | 3864 | 430 |
| 28 | 3990 | 4095 | 105 |
| 29 | 1242 | 1962 | 720 |
| 30 | 0 | 950 | 950 |
| 31 | 1222 | 2074 | 852 |
| 32 | 4095 | 3848 | -247 |
| 33 | 4095 | 3108 | -987 |
| 34 | 1222 | 1142 | -80 |
| 35 | 0 | 0 | 0 |
| 36 | 1242 | 2053 | 811 |
| 37 | 3990 | 3386 | -604 |

Similar to the previous cases, the channel values from the reconstruction resemble the initial distortion imposed by the deformable mirror.

We can conclude with a comparison between the algorithm of the plenoptic sensor and conventional Shack-Hartmann sensors shown in table 6:

Table 6. Comparison between plenoptic sensor and Shack-Hartmann sensors

| Properties | Plenoptic Sensor | Shack-Hartmann sensor |
|---|---|---|
| Plane of reconstruction | In the field | In front of the sensor |
| Geometrical information | Determined in each block | Determined by block index |
| Angular information | Determined by block index | Determined in each block |
| Separation of the light field | By patches inside the beam | By blocks in the MLA |
| Condition for interference patterns | If patches overlap geometrically and angularly at the same time | If patches overlaps geometrically |
| Reconstruction algorithm | Global reconstruction | Local reconstruction |

The first three properties in table 6 are demonstrated by the mechanism of the plenoptic sensor. An interesting observation in the comparison is the swap between geometric and angular information recovery in both wavefront sensors. Both types of sensors can record the 4D light field (2D in geometric and 2D in angles) by a 2D image. The difference is in extracting each light patch (or ray), the Shack-Hartmann sensor requires $n \times n$ pixels ("$n$" is determined by the width of a micro-lens divided by the width of a pixel). However, in the plenoptic sensor, each pixel corresponds to a light patch (or ray). Therefore, the plenoptic sensor can obtain light field information that is $n \times n$ times larger than the Shack-Hartmann sensor with the same pixel resolution. The fourth property refers to the fact that in the plenoptic sensor, the direction of a light patch is determined by the objective lens first, and then its geometric shape is revealed by the MLA. Thus if a beam breaks up into multiple patches, the patches are recorded depending on their own angular and geometric properties. On the contrary, the Shack-Hartmann sensor divides the incident beam geometrically by its sub-apertures. The fifth property states that for a plenoptic sensor, any two patches interfere with each other if and only if they are both geometrically and angularly overlapping. In the Shack-Hartmann sensor, if two patches arrive at the same MLA cell, two spots will show up in the same block. Then a logic problem of whether the spots originate from an abrupt bending in the wavefront or from two patches intercrossing each other arises in the reconstruction. The sixth property is due to the fact that the plenoptic sensor can be viewed as an array of Keplerian telescopes where each MLA cell provides information on a "global" scale. In other words, for any spot in the plane of reconstruction, the algorithm tries to find all its corresponding pixels in different blocks, and the risk of finding no information (if and only if all the corresponding pixels are not illuminated) will be significantly reduced. If a MLA cell in the Shack-Hartmann sensor is not illuminated, the area that corresponds to the plane of reconstruction can't be directly retrieved (but can be potentially interpolated if neighboring spots are recovered accurately). The wavefront reconstruction difficulty caused by the dark regions is commonly referred as a branch point problem[26]. The design of the plenoptic sensor is less affected by the branch point problem since the loss of illumination in one block can be generally found in other image blocks. Also, in situations where all the block images provide low intensity for a certain spot on the reconstruction plane, the algorithm can make a detour and recover the phase change. The detour process can be illustrated by figure 14:

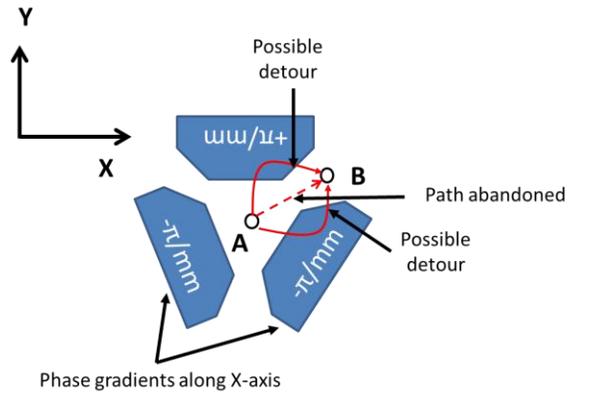

Figure 14: Detour process when the integral path fails to cover pixels with adequate intensity

For simplicity, we assume there is no phase gradient along the Y-axis in Figure 14. The colored islands are pixels that are illuminated while the remaining areas are "dark" with low intensity value. The last step of the reconstruction algorithm will find that the direct path from A-B lies in a dark area where the phase change is not correctly revealed. A detour process will be triggered by increasing the path length (number of pixels covered by the path) until the detoured path has enough illumination. Therefore, the "global" reconstruction algorithm of the plenoptic sensor is more robust against the branch point problems caused by low illumination. In practice, we find that the detour process is rarely triggered in our experiments.

### 2. Algorithm(2): Intelligent interferometer

The major source of error in algorithm (1) is that interference and diffraction are involved as the distorted beam propagates to the plenoptic sensor. Basic image filters, such as averaging over a small aperture, work effectively for basic modes of reconstruction. Since the smoothing function is a low band-pass filter of spatial frequencies, the reconstruction of basic Zernike modes is not significantly affected by interference or diffraction. However, to solve higher oscillation terms of phase changes, one needs to list the full image formation equations[15] and derive a solution with MMSE (minimum mean square error) estimation.

The approach in algorithm (2) is to place the deformation at the front focal plane of the objective lens. Then, diffraction and interference become negligible, and the plenoptic sensor works like an interferometer that can retrieve phase changes without using any reference beam.

To prove this point, we command the DM to enforce a symmetric Trefoil and use the plenoptic sensor to reconstruct the surface independently. Then, we compare the reconstruction result with the image obtained by an interferometer of the beam interfering with a collimated Gaussian beam[27]. The result is demonstrated in Figure 16. As a result, the contour map of the reconstruction matches the image from the interferometer.

The diagram of our interferometer is shown as:

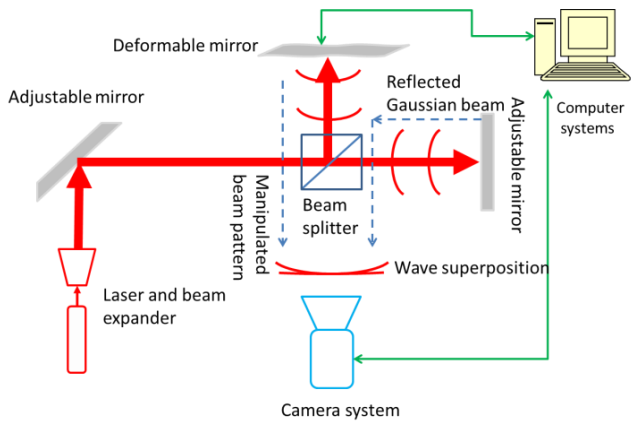

Figure 15: Diagram of interferometer used for comparison

The interferometer interprets the deformation's phase information with intensity variations on the image obtained[26]. The dark lines between the bright "islands" in the interferometer images mark the locations where the phase difference between the distorted beam and the reference beam is $(2n+1)\pi$ where n is an integer.

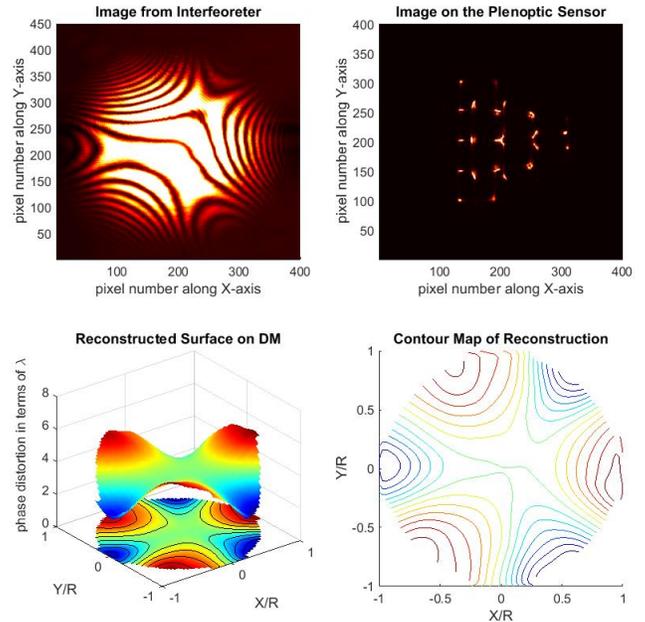

Figure 16: Using the plenoptic camera to achieve a similar result to an interferometer ("Trefoil" deformation is used)

In figure 16, the upper-left plot is the interferometer image and the upper-right plot is the image of the plenoptic sensor. The lower-left plot is the reconstruction results and the lower-right plot is the contour map for the reconstruction. The results show that the contour plot of the reconstruction agrees with the distribution of dark lines in the interferometer image. However, for the interferometer reconstruction, one has to deal with the logic errors for each dark line between different bright "islands" because bifurcated trends of phase changes can lead to the same result. In other words, 2 phase variations are possible that lead to the same intensity pattern. The logic error in interferometer reconstruction can be illustrated by figure 17:

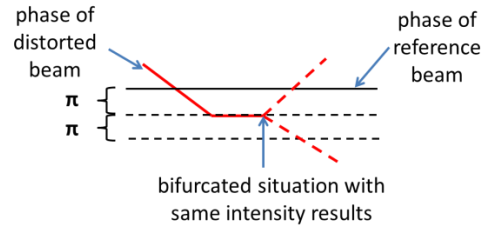

Figure 17: Illustration of logic errors in interferometers

The logic error can also be recognized in the interferometer image in figure 16. For any disjoint dark line, the trend of phase change at both sides of the line is unclear. Therefore, a reconstruction of the interferometer image is normally within range $(-\pi, \pi]$.

The plenoptic sensor can serve as an intelligent interferometer that doesn't require any reference beam. It doesn't have the same problem of logic errors in reconstruction. The underlying principle is that the plenoptic camera uses its lens system as a reference in the hardware layer. When a distorted beam is observed by the plenoptic sensor, the objective lens sorts the gradients of the phase distortion into different MLA cells and then the MLA cells can recover the geometric distribution for each gradient component respectively. Thus, on a laboratory scale, the plenoptic sensor can be used as an intelligent interferometer that can directly observe the distorted beam and retrieve the phase information ranged up to multiples of $2\pi$ ($8\pi$ for the "Trefoil" case in figure 16).

### 3. Algorithm(3): Multi phase screen reconstruction

We attempt to reconstruct more than one phase screen with the plenoptic sensor. The challenge is that only one plenoptic image can be used if there are two phase screens located at different depths in the channel. Intuitively, correct reconstruction on any one of the two phase screens will lead to the correct reconstruction of the other phase screen. In other words, if the latter phase screen is reconstructed correctly, one can retrieve the light field before it hits the second phase screen and the problem is simplified into algorithm (1): single phase screen reconstruction. Similarly the correct recovery of the former phase screen will lead to the correct reconstruction of the latter one.

In our experiment, we command a Trefoil on the DM's surface and place a randomized glass plate after the DM and in the beam's propagation path. The results are shown in Figure 18.

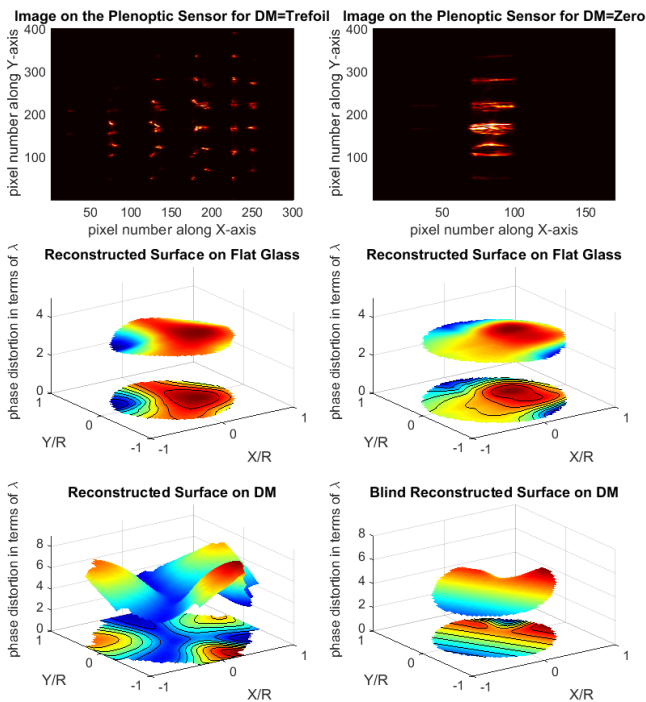

Figure 18: 2 phase screens reconstruction of a DM and a piece of unknown glass serving as an additional phase screen

In figure 18, the upper-left plot is the actual plenoptic image used for the reconstruction of the 2 phase screens. The upper-right plot is the plenoptic image of the randomized glass plate when the DM is set zero (flat mirror). Without loss of generality, one can reconstruct the phase distortion of the randomized glass plate by the single phase screen algorithm and set it as the reference for the algorithm of 2 phase screens reconstruction. The middle-left image is the reconstructed randomized glass plate when the DM is deformed with a "Trefoil" shape. The middle-right plot is the reconstructed result of the randomized glass plate when the DM is set to zero. The lower-left plot is the reconstructed DM surface in the 2 phase screen reconstruction. The lower-right plot is a blind reconstruction in the middle of the channel regardless of the fact that there are actually 2 phase screens.

According to the steps of algorithm (3), the largest patches in the plenoptic image (marked as set *{U}*) are selected and back propagated. Then, the plane $Z=z_1$, where the patches in set *{U}* combine most uniformly, is set as the first reconstruction plane (for the first event). The first phase screen at $Z=z_1$ is reconstructed using those patches in *{U}* that combine most uniformly. Based on the first reconstructed phase screen, the second phase screen can be reconstructed similarly except for one additional restriction: if $z<z_1$ in the back tracing of rays, each patch should subtract the phase change caused by the reconstructed phase screen. To evaluate the results of the 2 phase screen reconstruction, one can continue to use the correlation between the DM channels and reconstructed values on the channels for one of the phase screens. While for the other phase screen (the randomized glass plate), the correlation is calculated between the result of the 2 phase screen reconstruction and the result of the reconstruction when the DM is set flat (1 phase screen reconstruction). The correlation for the reconstruction on the DM is 0.8042, and the correlation for reconstructed randomized glass plate is 0.8141. The RMS error on the DM channels is $0.85\lambda$ (compared to maximum distortion of $5.5\lambda$). The correlation numbers support the claim that if one phase screen is reconstructed correctly, the other one will be mutually correct. Thus the loss of correctness should be mutual for the 2 reconstructed phase screens with one plenoptic image.

Blind reconstruction will lead to error because the randomized glass plate can diverge the light patches. Consequently, in the back tracing of light patches, the geometric information of the patches can't be correct. The result of this blind reconstruction also suggests that for some cases a single phase screen is not adequate to express the progressive effects of atmospheric turbulence in a channel.

Compared with the single phase screen reconstruction algorithm, the multi-phase screens reconstruction algorithm will have lower accuracy in each phase screen. The multi-phase screen reconstruction algorithm has a property that errors in any one of the phase screen reconstructions will mutually affect the accuracies of other phase screen's reconstructions. As a result, with more and more phase screens to be reconstructed with one plenoptic image, the distributed reconstruction algorithm will deteriorate in both efficiency and accuracy[28,29].

### 5. Conclusions

In this paper we demonstrated the structure, mechanism and algorithms of our plenoptic sensor as a new type of wavefront sensor to retrieve basic phase distortions of a beam. The amplitude distortions of a beam can be retrieved directly by the pixel illuminations. The major purpose of the reconstruction is to instruct AO systems to correct a distorted beam more intelligently. Since the single phase screen reconstruction algorithm is a linear superposition of the algorithm on each block image, it is compatible for parallel computing which can further reduce the processing time. The multi-phase screen algorithm demonstrates how to use one plenoptic image to reconstruct more than one phase screen in the turbulence channel. Therefore, a 3D reconstruction of the turbulence channel can be potentially achieved by this new type of wavefront sensor. However, the multi-phase screen reconstruction algorithm doesn't benefit from parallel computing as much as the single phase screen reconstruction does due to the non-uniform treatment for each block image[30].